**TECHNICAL REVIEW**

# Antennas for low-frequency radio telescope of SKA

AGARAM RAGHUNATHAN[1,*] , KEERTHIPRIYA SATISH[1], ARASI SATHYAMURTHY[1], T. PRABU[1], B. S. GIRISH[1], K. S. SRIVANI[1] and SHIV K. SETHI[2]

[1]Electronics Engineering Group, Raman Research Institute, Bangalore 560080, India.
[2]Astronomy and Astrophysics Group, Raman Research Institute, Bangalore 560080, India.
*Corresponding author. E-mail: raghu@rri.res.in

**Abstract.** The low-frequency radio telescope of the Square Kilometre Array (SKA) is being built by the international radio astronomical community to (i) have orders of magnitude higher sensitivity and (ii) be able to map the sky several hundred times faster, than any other existing facilities over the frequency range of 50–350 MHz. The sensitivity of a radio telescope array is in general, dependent upon the number of electromagnetic sensors used to receive the sky signal. The total number of them is further constrained by the effects of mutual coupling between the sensor elements, allowable grating lobes in their radiation patterns, etc. The operating frequency band is governed by the desired spatial and spectral responses, acceptable sidelobe and backlobe levels, radiation efficiency, polarization purity and calibratability of sensors' response. This paper presents a brief review of several broadband antennas considered as potential candidates by various engineering groups across the globe, for the low-frequency radio telescope of SKA covering the frequency range of 50–350 MHz, on the basis of their suitability for conducting primary scientific objectives.

## 1. Introduction

The low-frequency radio telescope of SKA (SKA-Low) is the world's largest low-frequency radio telescope being built to explore the Universe at wide range of frequencies starting from 50 MHz to 350 MHz. It is a global effort by several radio astronomical/engineering communities in different countries spread across the globe. The SKA-observatory is taking a leading role in this effort. The final SKA telescope is expected to be the most sensitive (of the order of 50 times (Schilizzi *et al.* 2008) compared to any other existing telescope in the world with a large instantaneous bandwidth of 300 MHz.

The major science programs that SKA-Low are expected to drive, will fall under four broad categories: (i) Studying the early Universe, which includes epoch of reionization (red shift *z* range: 15–6) , cosmic dawn (*z*: 30–15) and possibly later phases of dark age (*z*: 200–30) (Koopmans *et al.* 2015), (ii) understanding the role of dark matter on the evolution of Universe (Tingay & Hall 2012), (iii) pulsar survey and timing (Smits *et al.* 2009) and (iv) discovering rare objects like binary pulsar in galactic center, rotating radio transients (McLaughlin *et al.* 2006), etc., that act as greatest physics laboratories.

In the science case of studying the early universe, direct imaging of neutral hydrogen will be carried out over scales of arc-minutes to degrees over most of the redshift range *z*: 6–28 with SKA-Low. In pulsar timing and survey, majority of the pulsar population in the galaxy among binary, millisecond pulsars, transient and intermittent sources will be found out.

A low-frequency aperture array (LFAA) of wide-bandwidth wire antennas forms the low-frequency radio telescope of SKA (SKA-Low). The SKA-Low has 512 stations, each containing 256 dual linearly polarized antennas. The antennas in each station are phase combined to produce multiple beams in the sky.

The effective collecting area, antenna impedance, smoothness of the spectral response and cross polar performance are considered as important figures of merit for an antenna for its application in radio astronomy. The effective collecting area determines the directivity of the antenna. If the magnitudes of both these parameters, which are interdependent are constrained independently (refer Table 1 for SKA-Low requirement), then the structure may require modification to meet both the requirements simultaneously. The antenna impedance plays a major role in the noise performance of a radio receiver system. The requirement on its value is stringent at high frequencies (>100 MHz) since at those frequencies, the total system noise is determined mainly by the receiver electronics rather than sky. Constrained by the SKA requirement of noise figure (<1 dB over 50–100 MHz and <0.5 dB over 100–250 MHz), the structural dimensions of the antenna are optimized to get the desired antenna impedance. The antenna impedance is known to get influenced by the mutual coupling effects in an array of antennas. However, random placement of them within the array is expected to minimize this effect significantly. Features in the spectral response of the antenna are known to negatively impact the ability of the antenna to detect EoR and cosmic dawn signals. These features can arise due to either (i) impedance mismatch between antenna and low-noise amplifier in the signal path or (iii) due to chromatic behavior of the antenna itself. The magnitude of these features are desired to be kept minimum in the design process to meet the SKA requirement of spectral smoothness in both their magnitude and phase as indicated in Table 1. The cross polar performance in a dual polarized antenna is expected a play a very important role in the study of pulsar and its timing. This parameter quantifies leakage of electric signal between two orthogonally oriented antennas because of electromagnetic coupling of electric field between them. As per the SKA requirement, cross polar performance of atleast 11–12 dB should be ensured over the entire frequency band of 50–350 MHz.

The SKA-Low is expected to have (i) good point source sensitivity, (ii) enhanced surface brightness temperature sensitivity, (iii) large operating band, (iv) good cross polar performance and (v) large collecting area. The present paper reviews antennas proposed for the low-frequency radio telescope of SKA on the basis of several figures of merit like (i) effective collecting area, (ii) spectral/spatial response, (iii) mutual coupling between antennas and (vi) polarization purity. Section 2 provides review of different antennas proposed for SKA-Low. Section 3 talks about merits and demerits of the SKALA4.1 antenna selected for the low-frequency radio telescope of SKA. Section 4 discusses the impact of fine-scale structures in the bandpass characteristics of different SKA log-periodic antenna (SKALA) designs on EoR science. Conclusions of the paper are made in Section 5.

## 2. Review of antennas proposed for SKA-Low

Antennas proposed as potential candidates for the SKA-Low instrument by various groups across the world are briefly reviewed in this section. The characteristics of each one of them will be compared with the technical requirement needed (Bolli *et al.* 2020) as indicated in Table 1, to meet the science goals defined by the astronomy community for SKA.

### 2.1 *Vivaldi tapered slot antennas and conical antennas*

An aperture array based on Vivaldi tapered slot antennas was proposed by Schaubert *et al.* (2003) as a via solution for SKA-Low of SKA. A thousand element array (THEA) demonstrator built using these antennas indicated that it suffered from effects like truncation effect, trough resonance, higher cross coupling and non-smooth reflection coefficient. Due to these limitations, this antenna was found not suitable for SKA-Low application. The International Centre for Radio Astronomy (ICRAR) explored conical spiral antennas to propose it as one of the candidate antenna design for SKA-Low (Jiwani *et al.* 2011). Even though it had smooth patterns, it had unacceptable backlobes at low frequencies and beam tilts at higher frequencies. Due to these, this antenna was not considered for SKA-Low instrument. Further Jiwani *et al.* (2012) evaluated the wire version of conical antenna, which had however, unacceptable undulations in gain at 3 dB level and poor sidelobe response.

### 2.2 *SKA log-periodic array antenna (SKALA)*

SKA-Low instrument has highly demanding requirements like (i) wide band of operation (50–350 MHz), (ii) maximum sensitivity (A/Tsys) over field of view of ±45°, (iii) smooth spectral and spatial response, (iv) good impedance match, (v) polarization purity, (vi) cost and (vii) longevity, etc. as listed in Table 1. Given these, challenge is to design a dual polarization element capable of performing electromagnetically over the full frequency range maintaining high beamwidth and high polarization isolation.

Log periodic array antenna was proposed by de Lera Acedo *et al.* (2015b), as a candidate for the SKA-Low instrument, since it meets the requirements and provides basis for further development. The design challenges

**Table 1.** Specifications of SKA-Low instrument.

| Sl. no. | Name of parameter | Required value |
| --- | --- | --- |
| 1 | Frequency range | 50–350 MHz |
| 2 | Polarization | Dual linear |
| 3 | Directivity | 8 dBi |
| 4 | Directivity degradation at FoV (±45°) edge | Drop by 3 dB w.r.t zenith |
| 5 | Spectral smoothness of antenna response (mag.) | <0.02* |
| 6 | Spectral smoothness of antenna response (phase) | ±0.2°** |
| 7 | Static intrinsic cross polarization (50–250 MHz) | ≥12 dB |
| 8 | Static intrinsic cross polarization (>250 MHz) | ≥11 dB |
| 9 | Low Noise Amplifier (LNA) gain | ≥40 dB |
| 10 | LNA noise figure (50–100 MHz) | <1 dB |
| 11 | LNA noise figure (100–350 MHz) | <0.5 dB |
| 12 | Antenna foot print | 1.6m × 1.6m |

*Normalized fractional residual on voltage gain amplitude over 2.25 MHz band. **Phase gradient over 4.6 KHz.

were addressed in the following manner. Sensitivity was maintained high by going for (i) slightly directive antenna to maximize the gain within the field of view of ±45° and (ii) low-noise amplifier with less noise figure at higher frequency band (since sky is less bright at higher frequencies). To achieve still higher sensitivity, the filling factor of the SKA core is required to be high. As per SKA requirement, this is done by distributing 256 antennas within one station of 35 m diameter. However, this resulted in (i) higher level of mutual coupling between antenna elements, (ii) reduced foot-print area per element (1.93 × 1.93 sq m) and (iii) poor impedance mismatch at low frequency because of reduced foot print. The mutual coupling effects were minimized by randomly distributing the antennas within the station and the impedance mismatch at lower frequencies was improved by transforming bottom dipole into a bow-tie type radiating structure.

With these considerations, log periodic dipole array called SKALA was designed and optimized to meet the desired requirements. The optimized structure had nine dipoles feeding to a differential LNA. Top dipoles were not made to resonate, but kept as dummies to preserve wideband performance. The design targeted a minimum separation of 1.5 m between the two antennas to enable configuring the antennas to redistribute main sidelobes into a sea of minor sidelobes and average out mutual coupling effects. The final structure optimized for its characteristics had a footprint of 1.2 m × 1.2 m with a height of 1.8 m. To minimize the effect of soil on the antenna performance and to maintain similar performance for all antennas in the array positioned at different locations, a metallic screen was used under the antenna element.

The authors indicate that the first design of array element was undertaken using a single element with an assumption that in a randomized array, the mutual coupling effects tend to randomize out and the array pattern could be represented to a first order by the multiplication of array factor and single element pattern.

The low-noise amplifier used was a pseudo-differential amplifier to achieve minimum possible noise temperature. Noise figure and gain were set at 0.4 dB and 40 dB at the higher end of the band, respectively. The two LNA outputs were connected to a wideband transformer for converting balanced output to single ended output. The antenna was designed to provide optimum impedance to the LNA to obtain minimum noise figure. PCB boards of both the LNAs were located at the top of the antenna.

The photograph of the antenna structure fabricated out of steel wires and tubes is shown in Figure 1. The structural components were chosen keeping in view of less weight and ease of assembly. The entire antenna weighs about 1.6 kg.

Figure 2 shows the measured antenna impedance characteristics along with the simulation results. The measurements and the simulation results are showing similar trend throughout the band. The match is observed to be bad at frequencies <100 MHz. This is attributed to the restricted dimension of the bottom dipole because of limited footprint area. The measured and simulated radiation patterns at 200, 300 and 450 MHz in both E and H planes are shown in Figure 3. Directivity in this frequency range is observed to have smooth spatial variation in the range of 6–7.5 dB from zenith to 45° off zenith. The radiation efficiency calculated for both the cases of soil and metallic screen below the antenna is shown in Figure 4. Poor efficiency at lower frequencies is due to the absorption of the signal by soil.

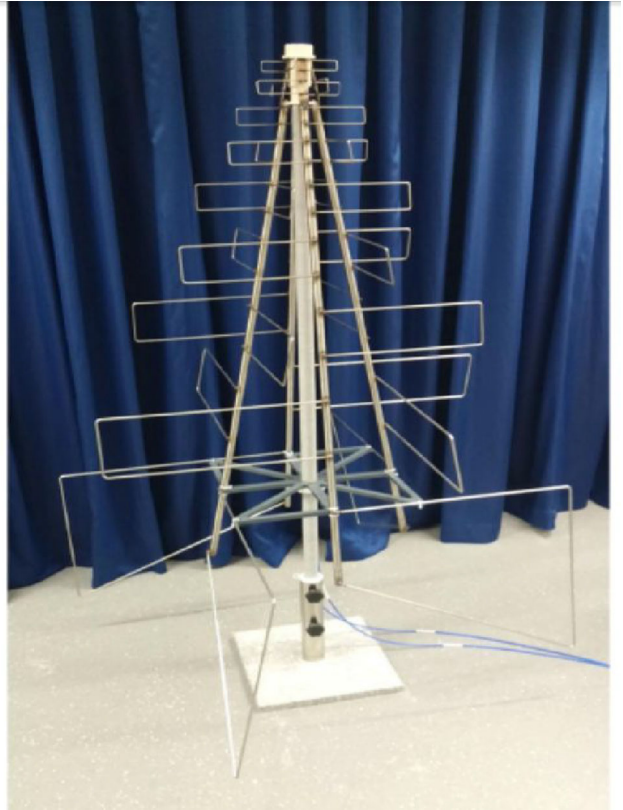

**Figure 1.** Photograph of fabricated SKALA antenna (de Lera Acedo *et al.* 2015b).

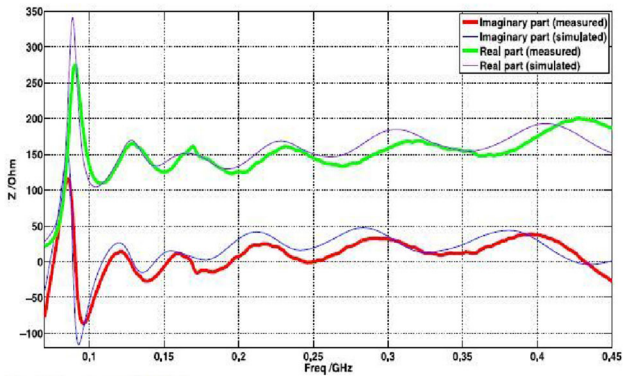

**Figure 2.** Input impedance characteristics of the SKALA antenna (de Lera Acedo *et al.* 2015b).

The gain of low-noise amplifier measured using the vector network analyser with hybrid splitter for feeding the differential input signals is shown in Figure 5. The gain is observed to be >40 dB over 75–400 MHz. The measured noise temperature of the differential amplifier when loaded with the antenna impedance is shown in Figure 6. The noise temperature is observed to be very high between 50 and 75 MHz owing to the poor impedance match as indicated in Figure 2. Knowing the gain of the antenna, receiver temperature of LNA and sky temperature, sensitivity was calculated as shown in Figure 7. It is observed to be more than 800 sq m per kelvin at zenith in the frequency range of 100–450 MHz.

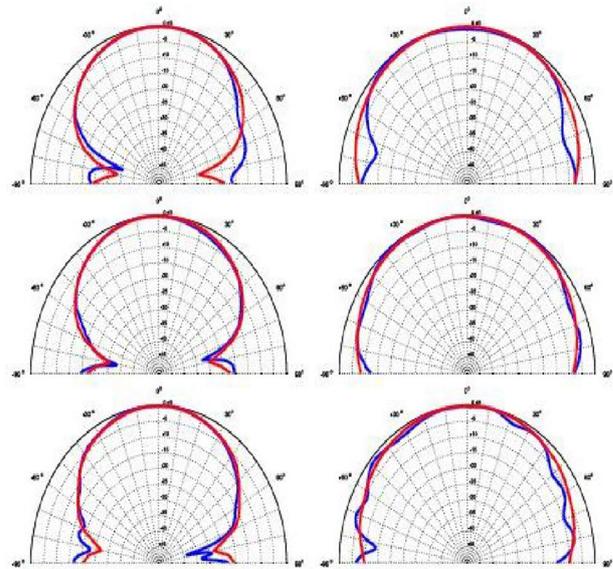

**Figure 3.** Measured (blue) and simulated (red) directivity patterns of SKALA on top of a 2 × 2 sq m ground plane at 200, 300 and 450 MHz (from top to bottom) for an E-plane cut (left 3 plots) and an H-plane cut (right 3 plots) (de Lera Acedo *et al.* 2015b).

A prototype array consisting of 3 × 3 SKALA antennas has been constructed to validate design and simulation results of antenna and low-noise amplifier. The cross coupling between adjacent antennas in the array has been measured and the results obtained are shown in Figure 8. The cross coupling of better than 20 dB has been achieved between adjacent elements in the array in the frequency range of 100–450 MHz.

The SKALA—the first version of antenna developed along with LNA is observed to have (i) poor noise performance (as high as 200–300 K instead of 75 K) and (ii) undesired spectral feature (gain kink at 3–4 dB level) in the frequency range of 50–100 MHz due to the impedance mismatch. In addition, it also had poor stability against wind load of 160 km per hour. The former two were attributed to the poor impedance match of the bottom dipole antenna, which is constrained in its dimension by the maximum foot print allowed as per SKA requirement. These limitations were addressed and improved in the second version of SKALA: SKALA2.

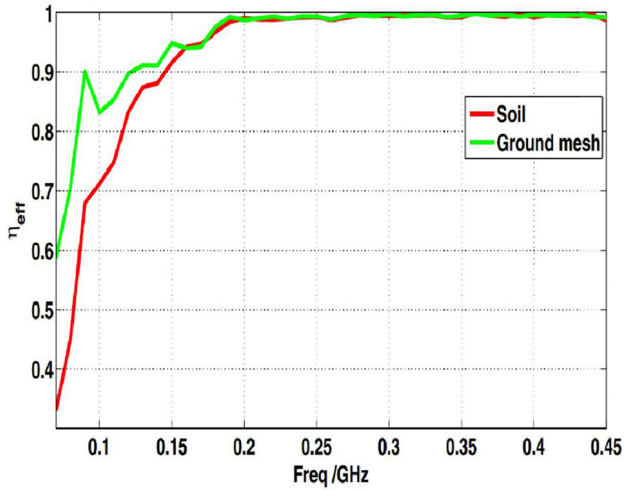

**Figure 4.** Computed radiation efficiency of SKALA antenna under different conditions (de Lera Acedo *et al.* 2015b).

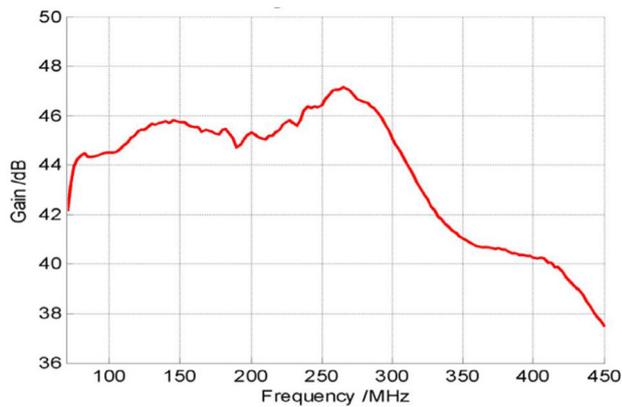

**Figure 5.** The measured gain of the low-noise amplifier (de Lera Acedo *et al.* 2015b).

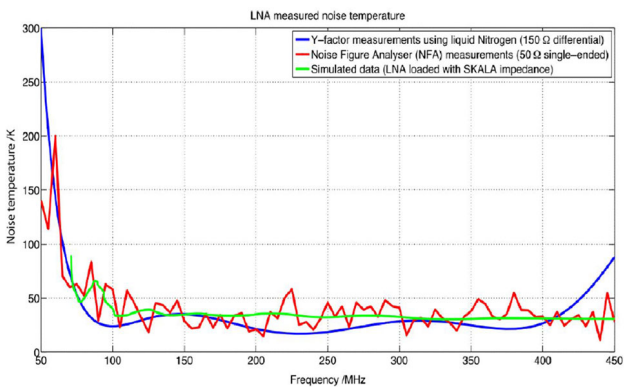

**Figure 6.** Noise temperature of differential amplifier when it is loaded with SKALA-1 antenna impedance. Blue and red curves represent measurements and green is the simulation. The noise temperature is <50 K over the band 75–450 MHz (de Lera Acedo *et al.* 2015b).

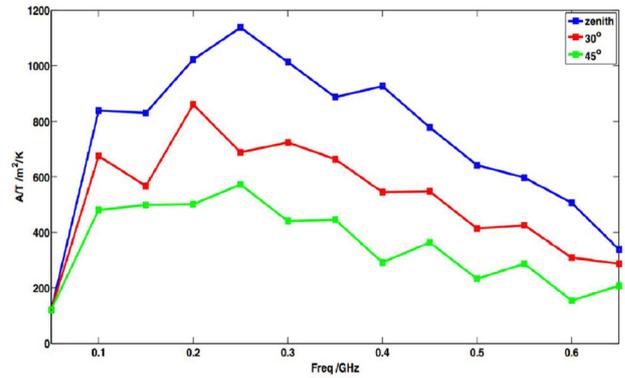

**Figure 7.** Sensitivity of SKALA-1 based on SKALA elements. Tsys (in K) is 4041, 817, 384, 210, 140, 106, 87, 76, 69, 64, 61, 58 and 61 at frequencies from 50 to 650 MHz in 50 MHz intervals (de Lera Acedo *et al.* 2015b).

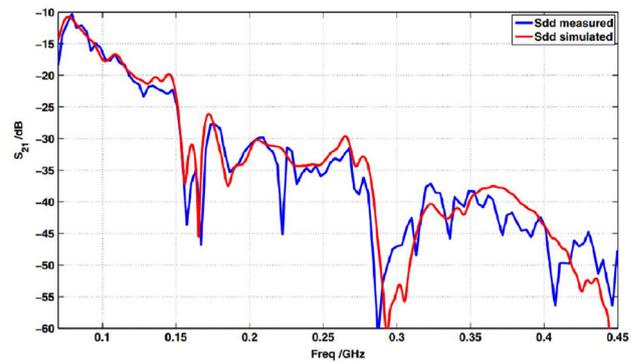

**Figure 8.** Coupling between central element and side element in a 3 × 3 regular configuration (de Lera Acedo *et al.* 2015b).

### 2.3 *SKALA-2*

SKALA-2 represents the upgraded version of SKALA (de Lera Acedo *et al.* 2015a) with more emphasis given to ease of manufacturability and deployment at a lower cost, while not sacrificing the performance characteristics of version 1. It also addresses robust support mechanism to meet the requirement of achieving vertical alignment error of less than ±1° under the wind speed of 160 km h$^{-1}$.

To meet the latter requirement, a heavy concrete base weighing about 70–100 kg was chosen and antenna was fixed to it using a bolt and nut arrangement. The field deployment of the antenna with this arrangement involved: (i) identifying the ground for antenna placement, (ii) rolling out the mesh that serves as a reflector, (iii) placement of pre-fabricated antenna base and adjusting its level using shims and (iv) fixing the antenna to the base. In this methodology, ground leveling of the entire station is not required. Instead, if leveling is done locally, that would be adequate. This simplifies significantly the deployment procedure in the field. The antenna's wire based construction ensured a lower cost per unit. The materials were chosen appropriately for the desert condition. The simulated structure of the SKALA-2 antenna is shown in Figure 9.

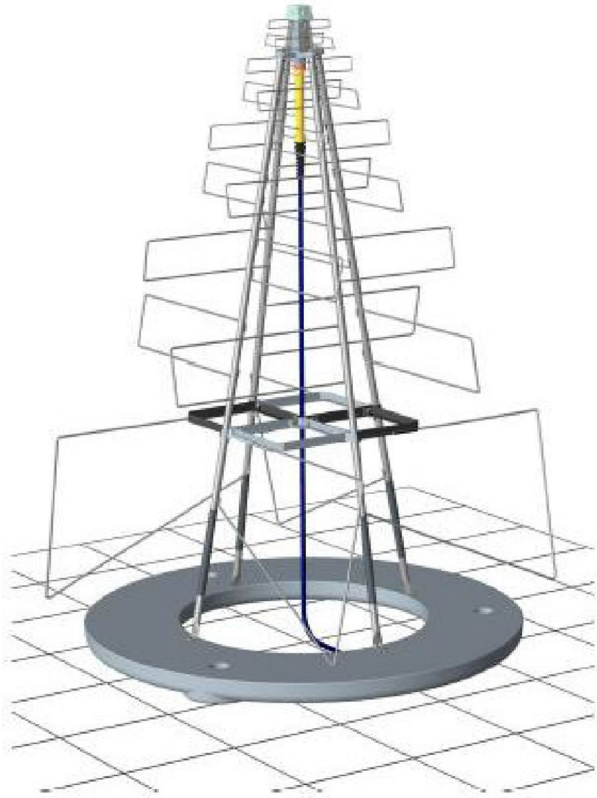

**Figure 9.** Computer model of SKALA-2 antenna (de Lera Acedo *et al.* 2017b).

In this version, LNA is redesigned with an emphasis on reducing the power consumption to <500 mW per polarization, while maintaining the gain and noise performance (40 dB and <35 K) of version 1. Qorvo TQP3M9039—a dual GaAs E-pHEMT LNA was used in the first stage of the amplifier. The connectors to the antenna terminals were upgraded to better resist the vibrations due to strong winds. The gain and noise performance of the upgraded LNA are shown in Figures 10 and 11, respectively, when connected to SKALA-2 antenna.

The spectral analysis of the bandshape of the LNA was performed to estimate the smoothness of the pass band response. A third order polynomial was fitted to it to obtain the normalized residuals. The magnitude of the normalized residuals is shown in Figure 14 and the phase gradient per fine channel (4.6 KHz wide) in Figure 15. It is observed from the plots that the required residuals (Trott & Wayth 2016) in magnitude are not satisfied by SKALA-2 antenna over the frequency range of 50–100 MHz. The magnitude of the observed normalized residual is about 0.04, which is two times higher than the desired value of 0.02 at around 60 MHz. However, the phase gradient requirement was under control than amplitude in the band 100–200 MHz, even though it is higher at lower frequencies. The improvement in the spectral smoothness to meet the desired specifications was carried out in the version 3 of SKALA: SKALA-3.

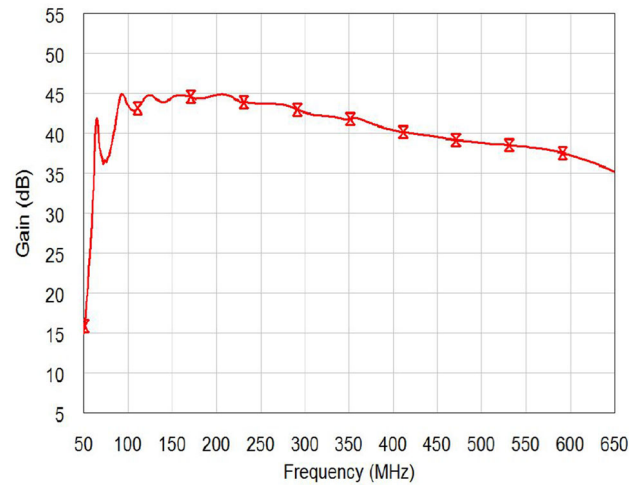

**Figure 10.** Gain characteristics of the version 2 LNA (de Lera Acedo *et al.* 2015a).

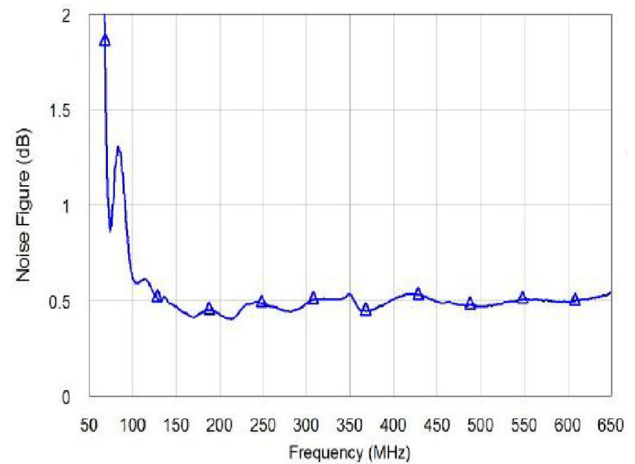

**Figure 11.** Noise characteristics of version 2 LNA when connected to SKALA-2 antenna (de Lera Acedo *et al.* 2015a).

### 2.4 *SKALA-3*

The limitation in SKALA-2 antenna in its spectral smoothness was addressed in SKALA-3 version (de Lera Acedo *et al.* 2017b) of log periodic antenna for low frequency instrument of SKA. Since the spectral response of the antenna is determined mainly by the impedance mismatch between itself and the first stage low-noise amplifier, both the antenna and the low-noise amplifier were modified to improve the spectral smoothness particularly, over the frequency range (50–100 MHz) where the desired pass-band requirements are not satisfied.

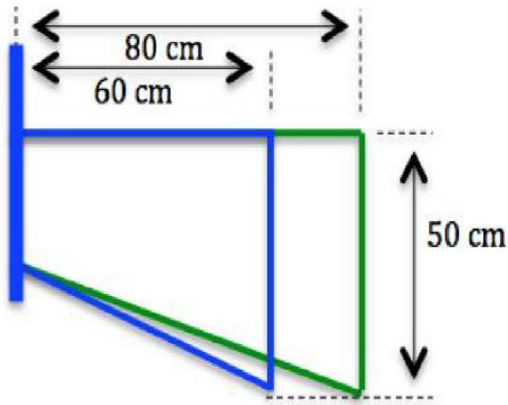

**Figure 12.** New lower antenna arm (green line, SKALA-3) versus old arm (blue line, SKALA-2) (de Lera Acedo *et al.* 2015b).

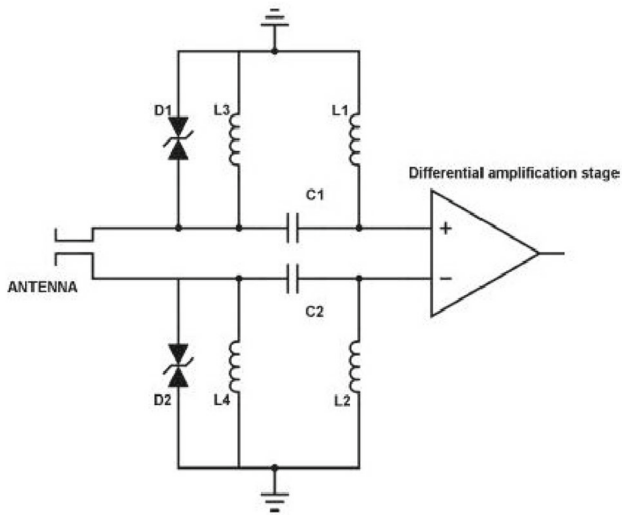

**Figure 13.** Simplified schematic of the first amplification stage in SKALA-3 antenna (de Lera Acedo *et al.* 2015b).

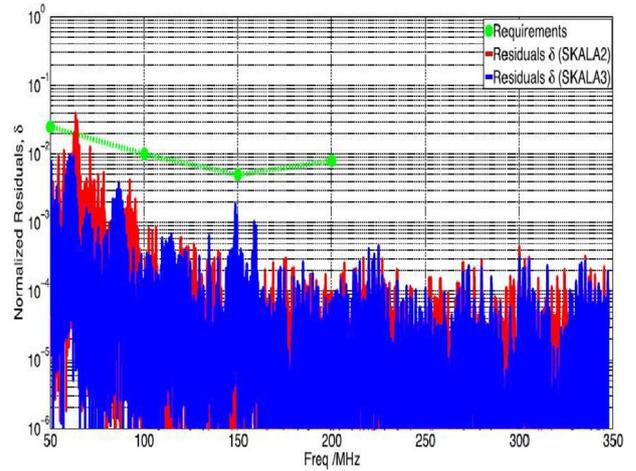

**Figure 14.** Normalized residuals obtained after version 2 (SKALA-2) and version 3 (SKALA-3) LNA band pass responses are fitted using a third order polynomial. Green curve represents the requirement (Trott & Wayth 2016).

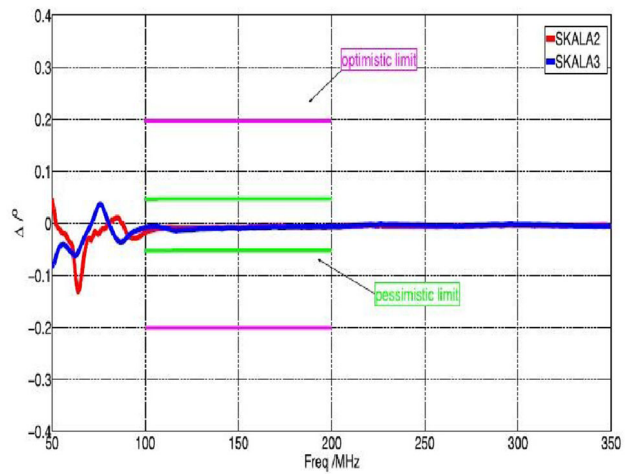

**Figure 15.** Phase gradient per fine channel of SKALA-2 and SKALA-3. The figure also shows the optimistic and pessimistic limits (Trott & Wayth 2016).

To have the antenna impedance (for the power match) closer to the LNA input impedance at low frequencies, the bottom dipole of the antenna was enlarged as shown in Figure 12. This modification resulted in an increase in the foot print from 1.2 m × 1.2 m to 1.6 m × 1.6 m.

The input matching circuit of LNA was modified to improve its power match with the antenna over the frequency range of 50–100 MHz. The simplified schematic of the input matching network is shown in Figure 13.

In the new design, lower values of DC coupling capacitors and bias inductors were used to improve the power match.

Both these changes resulted in the normalized residual magnitude and phase gradient within the desired limits (refer green curve) as shown in Figures 14 and 15. The green curves in these figures represent the maximum residual magnitude and phase tolerable after fitting the band pass response with a third order polynomial, to detect the EoR/CD signal in the red shift range of 25–6.1. Since this range of red shift corresponds to the frequency range of 50–200 MHz, the tolerable residual has been computed only till 200 MHz. In addition to residual magnitude and phase lying within

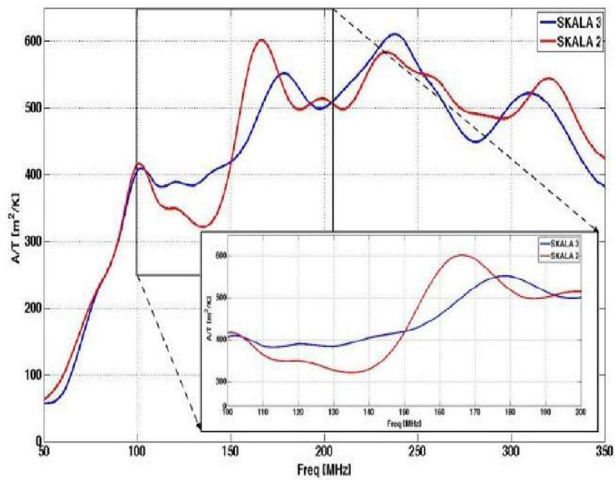

**Figure 16.** SKA1-Low sensitivity at zenith with SKALA-2 (red curve) and SKALA-3 (blue curve) across the full band (de Lera Acedo *et al.* 2017b).

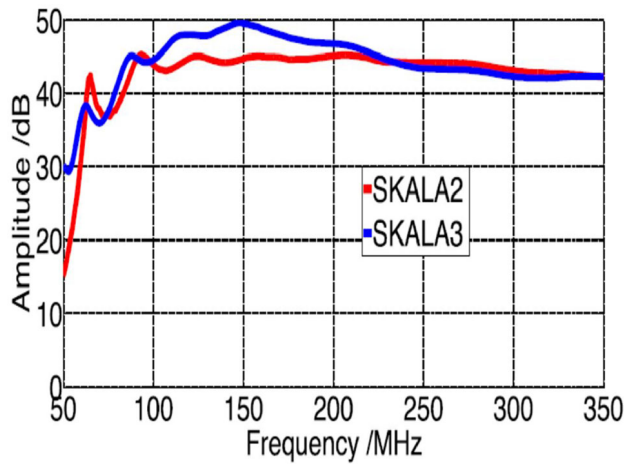

**Figure 17.** Amplitude of the LNA voltage passband when connected to SKALA-3 (blue) versus when connected to SKALA-2 (red). The plot shows the sharp feature found in the passband of SKALA-2 at 60 MHz (de Lera Acedo *et al.* 2017b).

limits, the sensitivity also became flatter across the frequency. The change in the dipole length is understood to have caused a smoother transition between bottom dipole and the second dipole resulting in a flatter directivity of the antenna across the frequency and hence, sensitivity as shown in Figure 16. Since impedance is matched for the maximum power transfer, the noise figure is observed to have increased to 8.6 dB at 57 MHz. The gain of the LNA when connected to the antenna is shown in Figure 17 for both SKALA-2 and SKALA-3 versions. Even though version 3 appears to be better than SKALA-2, it has undesired sharp spectral features at low frequencies. This is addressed in the next version of SKALA: SKALA 4.1, along with several others like

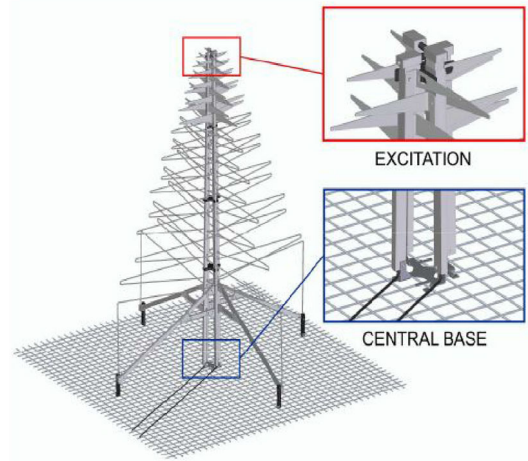

**Figure 18.** 3D CAD model of SKALA-4.1 antenna. The insets on the right show two crossed feeding lines on top of the antenna and at the bottom two coaxial cables carrying signal (Bolli *et al.* 2020).

improved spectral smoothness and improved directivity, etc., across the frequency band.

### 2.5 *SKALA-4.1*

In this version of antenna (Bolli *et al.* 2020), emphasis was given to (i) improve poor low frequency response as a result of severe constraint on the footprint of the bottom antenna, (ii) achieve large improvement in the spectral smoothness, (iii) achieve higher consistency between simulation and measurement and high degree of testability for array elements, (iv) improve directivity across frequency and front to back ratio, (v) protect LNA from electrostatic discharge and (vi) choose proper material to resist against weather.

The poor low-frequency performance due to constrained footprint of the bottom dipole was overcome by going for reinforced bow-tie shape for better impedance matching. The modified antenna with a change in the shape of the bottom dipole shown in Figure 18. Improvement in the gain response in the frequency range of 50–100 MHz shown in Figure 19 was achieved by tuning the LNA for its input impedance.

It was observed that the 50-ohm singled ended feeding for the antenna improved the spectral smoothness. This allowed a fruitful co-design between antenna and LNA. The two booms of the antenna which act as feed lines are connected to the LNA as shown in Figure 20 (de Lera Acedo *et al.* 2017a). The LNA output is taken through a shielded cable concealed completely inside the boom to the antenna base. This type of configuration did not require a balun since the shielded cable inside the boom cannot radiate outside the boom or couple to the dipole feeding lines. The single-ended configuration allowed both antenna and LNA characterization to be done with standard measurement equipment. The antenna booms were appropriately designed to have a 50 ohm output. Tubes of rectangular cross-section of 25 mm × 15 mm were chosen as booms with a gap of 23.5 mm at the top and 95.4 mm at the bottom. The computed antenna reflection coefficient along with the measurement is plotted in Figure 21 for the two polarizations in the frequency range of 40–360 MHz. The two ports corresponding to two polarizations are 1 cm apart along the vertical direction (see top inset of Figure 18).

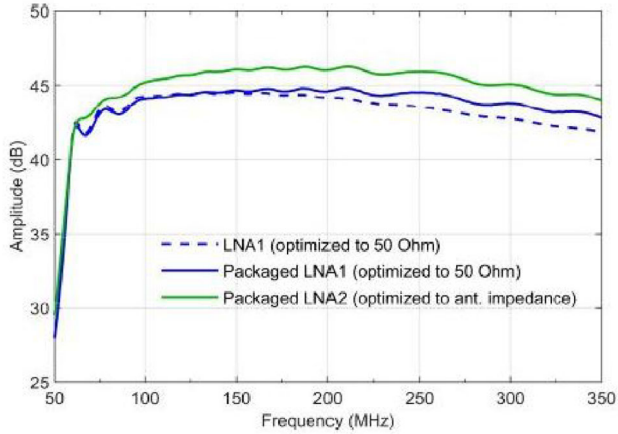

**Figure 19.** LNA gain response (SKALA4.1 antenna) under different input impedance conditions. Reduction in the ripple amplitude in the green curve is clearly seen, indicating improvement in the smoothness in the frequency range of 50–100 MHz (Bolli *et al.* 2020).

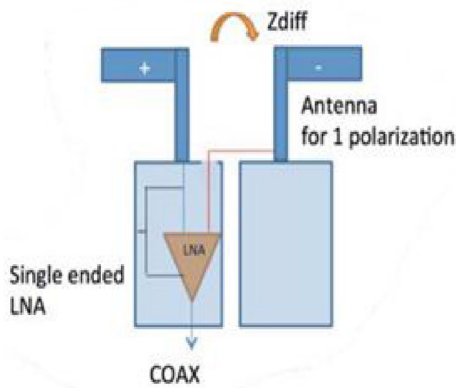

**Figure 20.** Shown in the figure are the two arms of a dipole connected to the single ended LNA via booms (de Lera Acedo *et al.* 2017a).

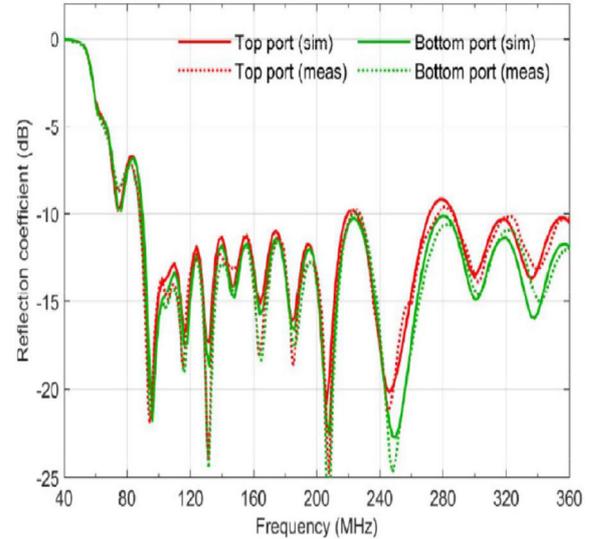

**Figure 21.** Simulated (continuous line) and measured (dashed line) reflection coefficient for the two SKALA4.1 antenna ports in the frequency range of 40–360 MHz. The reference impedance is 50 ohm (Bolli *et al.* 2020).

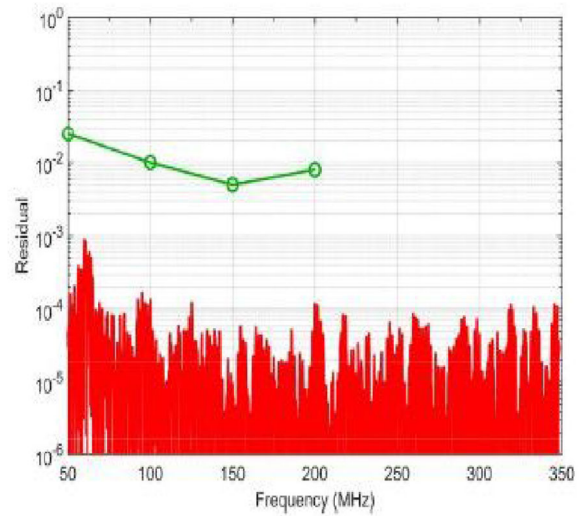

**Figure 22.** Spectral smoothness performance. Red curve shows computed and green curve shows the scientific requirement (Bolli *et al.* 2020).

The spectral smoothness achieved by adopting single ended 50 ohm feeding system and modified boom configuration is shown in Figure 22. The normalized residual magnitude computed was found more than an order of magnitude lower than the requirement.

The configuration of booms adopted to have 50 ohm output was observed to produce glitches in the directivity frequency response of the antenna as shown in Figure 23. This effect was mitigated by having a more number of dipoles and with an opening angle for the booms. Twenty dipoles and an opening angle of 1° between the booms were found to produce significant reduction in the glitches.

The mounting arrangement for the LNA at the top of the antenna and its interface resulted in having

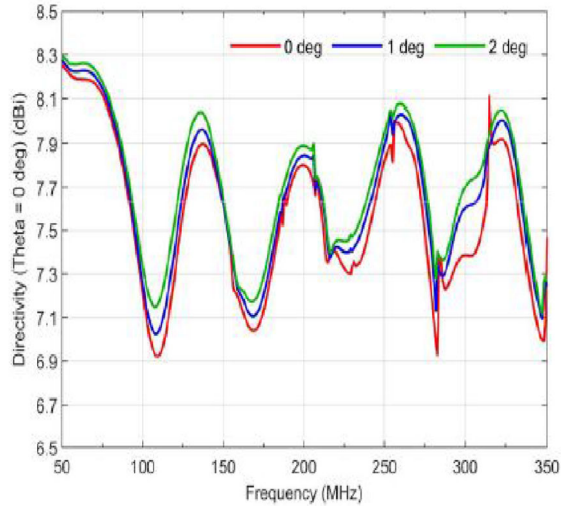

**Figure 23.** Directivity of the 50-ohm single-ended antenna for different values of the boom opening angles (0, 1 and 2°) at 0° zenith angle. The responses are observed to have glitches at different frequencies (Bolli *et al.* 2020).

(i) high consistency between simulation and measurement results and (ii) high degree of testability for array elements.

The improvement in both directivity and front to back ratio across the frequency is achieved by selecting triangular shape for the dipoles of the antenna. The number wired versus solid dipoles was decided based on the trade off between the directivity required and the resistance for the wind load. The simulated directivity of the antenna at zenith and 45° off zenith in both E and H polarizations are shown in Figure 24. The average directivity at zenith across frequency is 7.9 dBi with a maximum peak-to-peak variation of 1 dB. The slow ripple (about 60 MHz period) is determined by the antenna back-lobe from the ground plane towards the sky. The antenna is found to have a front-to-back ratio of >20 dB for frequencies above 100 MHz and >25 dB above 130 MHz (Bolli *et al.* 2020). The directivity in E plane drops > 8 dB at 45° from zenith, whereas in H plane, it drops <3 dB.

Typical embedded E-plane patterns at 50, 70, 160 and 320 MHz are shown in Figures 25 and 26. Plots indicate that the half power beam width in E plane varies from 74° to 56° over 50–320 MHz.

The radiation efficiency of the antenna was computed assuming that the electrical conductivity of aluminium is 1.4E7 S m$^{-1}$. It has a very flat behavior above 64 MHz (>99%), while it decreases to 90% at the lowest operating frequency (50 MHz).

The polarization purity performance is higher at zenith and deteriorates towards lower elevations. At high frequency, the cross polar values are <15 dB at low elevation angle. Only at 50 MHz, the cross polar value is always >15 dB in the entire field of view, while at 200 and 350 MHz, it matches the limits, i.e., 12 dB, between 50 and 250 MHz and 11 dB, above 250 MHz set by major science goals.

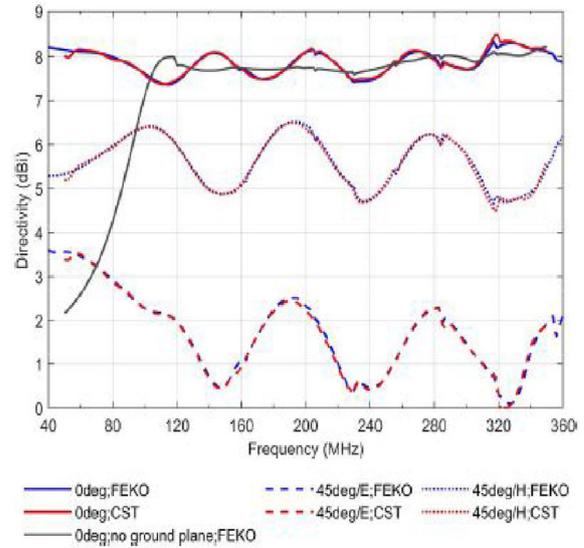

**Figure 24.** Directivity at zenith (solid curves) and at 45° off-zenith for both principle planes (dotted curves in H-plane and dashed curves in E-plane) in the frequency band of 40–360 MHz (Bolli *et al.* 2020).

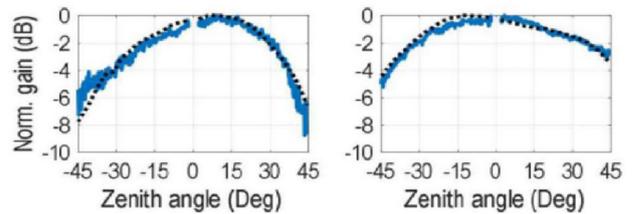

**Figure 25.** Normalized embedded element patterns of SKALA 4.1 antenna (y-polarization) at 50 and 70 MHz (Bolli *et al.* 2020).

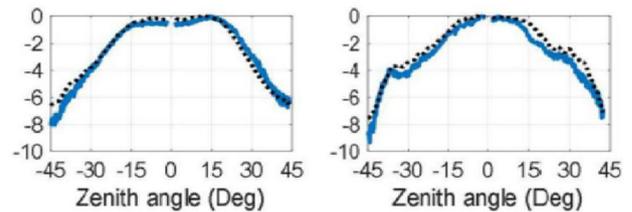

**Figure 26.** Normalized embedded element patterns of SKALA 4.1 antenna (y-polarization) at 160 and 320 MHz (Bolli *et al.* 2020).

The LNAs are protected against electrostatic discharge by short circuited grounding of the antenna. The anticorodal aluminum was used for the antenna for better electrical conductivity, lightness and cost when compared to other materials.

SKALA 4.1 antenna satisfies most of the SKA-Low instrument requirements. The average directivity at zenith across frequency is 7.9 dBi with a maximum peak-to-peak variation of 1 dB. The normalized residual magnitude computed in the band pass is order of magnitude lower than the requirement. The spatial gain variation requirement is not satisfied since E-plane directivity drops by >3 dB at the edge of field-of-view. The cross polar requirement is satisfied with 12 dB in the frequency range of 50–250 MHz and >11 dB beyond 250 MHz. Some of the responses which need improvement are (i) response at low frequencies, which has undesired spectral features and (ii) glitches in the directivity. The electrical characteristics of SKALA4.1 antenna are briefly described in the below section with some experimental verification and data modeling.

## 3. Electrical characteristics of SKALA4.1 antenna

The performance characteristics of SKALA4.1 antenna was evaluated both by modeling the data obtained through simulation and performing certain specific measurements to assess its characteristics like (i) spectral smoothness, (ii) glitches in the directivity, (iii) stability of the receiver as a function of time, (iv) cross talk between antennas and (v) effect of amplitude and phase errors on the station beam width and sidelobe levels. The results obtained have been documented by Ravi Subrahmanyan and others as memos in the link, https://www.atnf.csiro.au/observers/memos/. The evaluation of the above characteristics are briefly described in the sections below.

### 3.1 *Measurement of SKALA4.1 antenna band shape and receiver stability*

The total power spectra recorded over 50–350 MHz from Aperture Array Verification System Version 2 (AAVS2) over three days, were carefully examined and studied to understand its spectral shape and features in the frequency band of 50–350 MHz. The AAVS2 is an array of 256 SKALA4.1 antennas distributed in a semi-random manner over circular station area with a maximum spacing of 38 m. Along with this, time stability of the receiver system was also examined. All the spectra acquired were normalized to a reference power in the RFI free band of 65–80 MHz. The results of investigation are discussed in the following memo (https://www.atnf.csiro.au/observers/memos/The spectral response of embedded antennas in AAVS2.pdf).

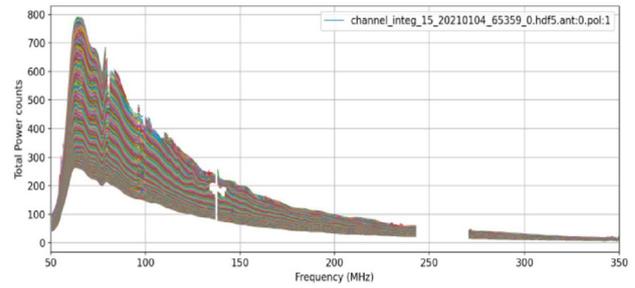

**Figure 27.** Measured spectra over 63 h overlaid. Wiggles in the spectrum are found stable with time (https://www.atnf.csiro.au/observers/memos/).

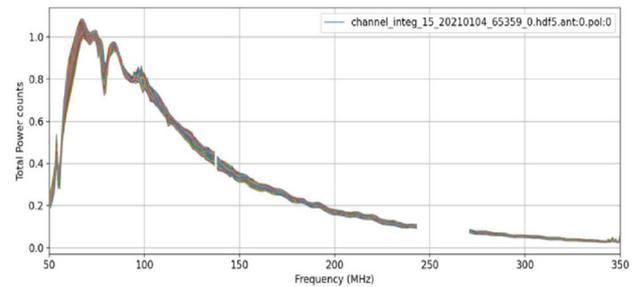

**Figure 28.** Measured spectra in X-polarization. Plot indicates significant undesired features at 55 and 78 MHz (https://www.atnf.csiro.au/observers/memos/).

Wiggles in the spectra were found reasonably stable with time. These are attributed to the property of the log periodic structure (Figure 27). Significant undesired spectral feature of considerable magnitude was found centered at 78 MHz in almost all the antennas (Figure 28). This feature is related to the average spacing between antennas in the station coupled with the phase change on scattering. It is at this frequency that the SKA-Low is targeting cosmic dawn detection. Many antennas also have sharp exhibited structure at 55 MHz. The spectral structures are found to differ between antennas and polarizations. The band shapes appeared uncorrelated between antennas of similar and cross polarization are shown in Figures 28 and 29.

To study the stability of the receiver, total power in the band of 65–80 MHz was plotted as a function of time. The dips in gain observed (Figure 30) over time are attributed to the RF over fiber (RFoF) modulator, which might have a feedback to stabilize the laser power. Unless calibrated, the gain dips will, however, cause change in the shapes of the beams formed by the stations, at times when significant RFI impacts a station. In the same way, all the band pass structures in the gain of the station beam versus frequency could be corrected for in the data as part of bandpass calibration. The bandpass calibration is suggested to be performed on a per-channel basis, which is the standard

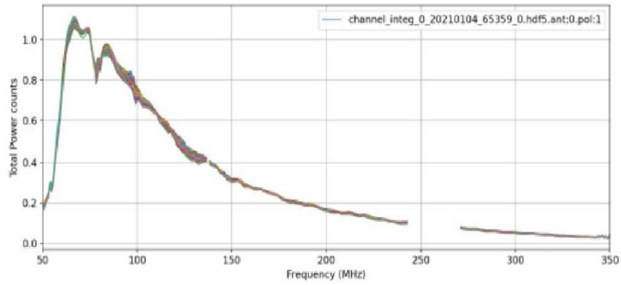

**Figure 29.** Measured spectra in Y polarizations (https://www.atnf.csiro.au/observers/memos/).

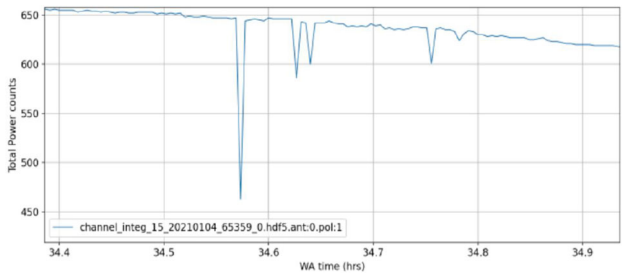

**Figure 30.** Measured receiver gain dips of varying magnitude with time (https://www.atnf.csiro.au/observers/memos/).

practice in Fourier synthesis telescopes like very large array (VLA).

### 3.2 *Measurement of fine scale frequency structures (glitches) in the frequency response of SKALA4.1 antenna*

Measurement of the antenna frequency response was made (refer memo: https://www.atnf.csiro.au/observers/memos/Anechoic chamber measurement of spectral glitches in SKALA4.1 antenna.pdf and Kyriakou *et al.* 2021) and to examine the fine spectral structure expected from the electromagnetic simulations. Measurements were calculated in the CIRA anechoic chamber at Curtin University. SKALA4.1 antenna and a standard commercial biconical antenna—BBAK 9137 along with VHBB 9124 balun (both from Schwarzbeck company) were used as transmit and receive antennas. A FieldFox network analyser was used to measure S21 parameter between them. Similar measurement was made between two biconical antennas for comparison. The measured S21 characteristics over a narrow frequency range of 200–214 MHz for both the above cases are shown in Figure 31.

The measurements indicate the existence of narrow band (0.5–1.0 MHz wide) structures in the S21 characteristics of biconical and SKALA4.1 pair of antennas and not in the biconical and biconical pair and thus confirming the presence of structures in the bandpass response of SKALA4.1 antenna with amplitudes ranging from 0.25 to 1.0 dB. The author of the memo is of the opinion that these features need to be quantified with a field measurement and using sky data. If the presence of feature is confirmed even in the sky data, the CD/EoR power spectrum observations might require bandpass calibration on a per-channel basis.

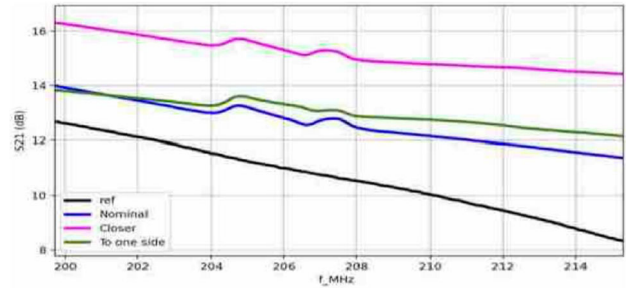

**Figure 31.** Measured S21 characteristics: Magenta, green and blue curves represent responses between SKALA4.1 antenna and standard biconical antenna with different distances between them. Black is the reference curve obtained between two biconical antennas (https://www.atnf.csiro.au/observers/memos/).

### 3.3 *Quantitative estimate of unwanted cross-talk between SKALA4.1 antennas at 110 MHz*

From the measurement results of noise parameters and S parameters of LNA, forward and reverse traveling noise powers were computed. Using the simulation results of scattering parameters (S11 and S21) of SKALA4.1 antenna configured in a station, amount of noise coupling was calculated at 110 MHz, which is closer to the geometric center frequency of the SKA-Low band. Cross-product of coupled noise voltage and the forward traveling noise voltage were computed to determine the magnitude of the cross-talk between two antennas (refer memo: https://www.atnf.csiro.au/observers/memos/Modelling the crosstalk between antennas in AAVS2.pdf). The magnitude of coupling resulting (Figure 32) in the cross-talk is observed to be <1 K between closely spaced antennas and <100 mK beyond 20 m intra-station baseline.

### 3.4 *Stability of the receiver system and robustness of station beam*

Total power data was acquired on sky over 24 h with X and Y polarization feeds of the SKALA4.1 antennas and analysis has been done at 110 MHz (refer memo: https://www.atnf.csiro.au/observers/memos/ Modelling the total powers of antennas in AAVS2 at 110 MHz.pdf) to measure the antenna temperature. The prediction of the antenna temperature made using global sky model and antenna embedded element patterns matched reasonably well with measurement data of 3–5% accuracy. The limitation in the accuracy was attributed to slow diurnal drifts in gain of the receiver chains, which is about 3 dB in X and 1.75 dB in Y polarizations and rapid gain fluctuations of about 0.1–0.2 dB during daylight hours. The author of the memo thinks that the cause of the drift requires investigation.

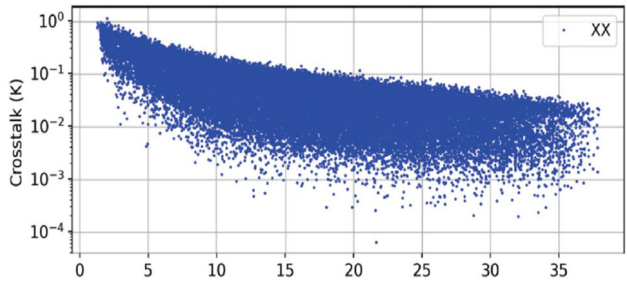

**Figure 32.** Magnitude of cross talk expressed in Kelvin for various intra-station baselines expressed in m (https://www.atnf.csiro.au/observers/memos/).

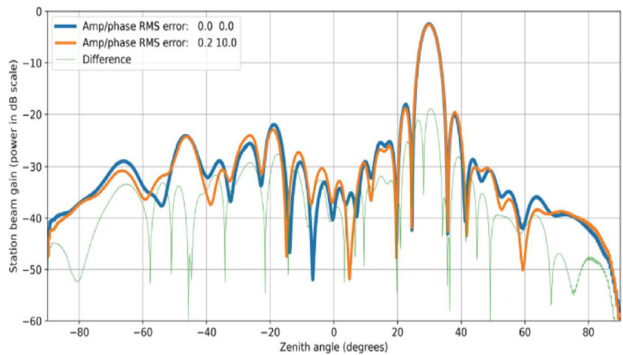

**Figure 33.** The station beam phased towards −135° azimuth angle and 30° zenith angle (https://www.atnf.csiro.au/observers/memos/).

The station beam patterns using SKALA4.1 antennas was computed (refer memo: https://www.atnf.csiro.au/observers/memos/Modelling the station beam of AAVS2 at 110 MHz.pdf) at 110 MHz for X polarization using embedded element patterns. The effect of antenna based amplitude/phase calibration errors on the quality of the station beam was examined. The analysis indicates that the station beam is fairly robust (Figure 33) to significant calibration errors particularly amplitude due to large number of elements. It has been found out that a 10° phase error together with 20% amplitude errors will perturb the beam main lobe and sidelobes at a level of 10%.

## 4. Discussion on impact of fine-scale structures in the bandpass characteristics of different SKALA antenna designs on EoR science

One of the main goals of SKA-Low instrument is to detect red-shifted 21 cm signal from the EoR and estimate the spatial properties of neutral hydrogen from the early Universe. The brightness temperature of this signal is predicted to be orders of magnitude weaker than the foregrounds and other instrumental systematics. Detection of this line can be achieved in two different ways: (i) by measuring the global signal averaged over the entire sky and (ii) by conducting statistical experiment such as measuring the power spectrum of the neutral hydrogen to estimate the spatial fluctuation of its brightness temperature. Global signal detection requires just a single antenna, whereas statistical measurement requires an array of antennas operating in the interferometric mode. The Square Kilometre Array is expected to measure the power spectrum over the red shift range of 6–25. This corresponds to the frequency range of ∼50–200 MHz.

Power spectrum is measured using the interferometer to determine the spatial distribution of the neutral hydrogen. The amount of fluctuation in the distribution is statistically measured by computing the variance in the measured power spectrum. These fluctuations are predicted to vary with time owing to the expansion of the Universe. The time evolution of these fluctuations is found by measuring power spectrum at different frequencies and computing the variance in each of them. Maps produced at different frequencies thus results in an image cube of neutral hydrogen depicting the distribution of neutral hydrogen both in space and time coordinates.

The detection of red-shifted 21 cm signal is primarily limited by the inadequate sky calibration model and the thermal noise level in the calibration data. Presence of foreground continuum sources (galactic and extragalactic) and the intrinsic chromatic response of the antenna also make the detection harder.

In the process of detection, foregrounds will be subtracted from the measured sky signal. The residuals thus obtained will have contributions from signals like thermal noise, spectral features and other artifacts, which are left out uncalibrated, along with the EoR signal. The contribution from the artifacts and features in the band pass are required to be smaller than the thermal noise power for successful detection of the EoR signal. To ensure this, limits were set by (Trott & Wayth 2016) on the magnitude of the spectral features and the phase gradient across the spectral channels. SKALA, SKALA2,

SKALA3 and SKALA4.1 were the improvised versions of log periodic antenna developed for SKA over several years. When the band pass response of each one of them was examined for its smoothness, SKALA3 and SKALA4.1 antennas were found to have adequate performance in terms of spectral smoothness for the detection of EoR signal, while SKALA and SKALA2 had residuals having power more than the thermal noise thus, impeding the detection of cosmological signal.

## 5. Conclusion

A brief review of broadband antennas considered as potential candidates by various engineering groups across the globe, for the low frequency instrument of SKA covering the frequency range of 50–350 MHz, has been made on the basis of their suitability for conducting primary scientific objectives. It includes vivaldi antenna, conical antenna and all the SKALA antennas from version 1 to version 4.1. While discussing the evolution of SKALA antennas, design details of each of the versions along with its performance characteristics has been presented. In each version, the modifications incorporated to improve the performance characteristics and the results obtained have also been briefly discussed. The merits and demerits of SKALA4.1 antenna chosen for SKA-Low instrument are briefly discussed based on certain well-defined experiments performed. The suggestions are also included to mitigate the effects due to the limitations observed in the antenna performance, before using it for scientific investigation.

## Acknowledgements

We acknowledge the help extended by the members of the Electronics Engineering Group at the Raman Research Institute in the various discussions we had, while writing this review paper.

## References

Bolli P., Mezzadrelli L., Monari J., *et al.* 2020, IEEE Open Journal of Antennas and Propagation, 1, 253

de Lera Acedo E., Drought N., Wakley B., Faulkner A. 2015a, in 2015 International Conference on Electromagnetics in Advanced Applications (ICEAA), IEEE, 839

de Lera Acedo E., Razavi-Ghods N., Troop N., Drought N., Faulkner A. 2015b, Experimental Astronomy, 39, 567

de Lera Acedo E., Virone G., Bolli P., *et al.* 2017a, LFAA Antenna & LNA Work Package, AADC Consortium, 17

de Lera Acedo E., Trott C. M., Wayth R. B., *et al.* 2017b, Monthly Notices of the Royal Astronomical Society, 469, 2662

Jiwani A., Juswardy B., Padhi S., Bij de Vaate J. G., Hall P. 2011, Active antenna development for the SKA, Asia-Pacific Microwave Conference 2011, Melbourne, VIC, Australia, p. 1186

Jiwani A., Padhi S., Waterson M., Hall P. J., Bij de Vaate J. G. 2012, Candidate wire spiral antennas for the SKA radio telescope, 2012 International Conference on Electromagnetics in Advanced Applications, Cape Town, South Africa, p. 510, https://doi.org/10.1109/ICEAA.2012.6328680

Koopmans L., Pritchard J., Mellema G., *et al.* 2015, arXiv preprint arXiv:1505.07568

Kyriakou G., Bolli P., Subrahmanyan R., Davidson D. B. 2021, in 2021 IEEE International Symposium on Antennas and Propagation and USNC-URSI Radio Science Meeting (APS/URSI), IEEE, 47

McLaughlin M. A., Lyne A., Lorimer D., *et al.* 2006, Nature, 439, 817

Schaubert D. H., Boryssenko A. O., van Ardenne A., Bij de Vaate J. G., Craeye C. 2003, The square kilometer array (SKA) antenna, IEEE International Symposium on Phased Array Systems and Technology, Boston, MA, USA, p. 351, https://doi.org/10.1109/PAST.2003.1257007

Schilizzi R. T., Dewdney P. E., Lazio T. J. W. 2008, in Ground-based and Airborne Telescopes II, Vol. 7012, SPIE, 603

Smits R., Kramer M., Stappers B., *et al.* 2009, Astronomy & Astrophysics, 493, 1161

Tingay S., Hall P. 2012, Australian Physics, 49, 174

Trott C. M., Wayth R. B. 2016, Publications of the Astronomical Society of Australia, 33, e019